\RequirePackage{fix-cm}
\documentclass[twocolumn]{svjour3}          
\usepackage[utf8]{inputenc}
\smartqed  
\usepackage{graphicx}
\usepackage{amsmath,amssymb,amsfonts}

\usepackage[printonlyused,nolist]{acronym}
\usepackage{enumerate}
\usepackage{paralist}
\usepackage{csquotes}
\usepackage{flushend}

\makeatletter
\let\cl@chapter\undefined
\makeatletter
\usepackage[capitalise]{cleveref}
\usepackage{url}
\usepackage{xcolor}
\usepackage{listings}

\colorlet{punct}{red!60!black}
\definecolor{background}{HTML}{EEEEEE}
\definecolor{delim}{RGB}{20,105,176}
\colorlet{numb}{magenta!60!black}

\lstdefinelanguage{json}{
    basicstyle=\normalfont\ttfamily,
    numbers=left,
    numberstyle=\scriptsize,
    stepnumber=1,
    numbersep=8pt,
    showstringspaces=false,
    breaklines=true,
    frame=lines,
    backgroundcolor=\color{background},
    literate=
     *{0}{{{\color{numb}0}}}{1}
      {1}{{{\color{numb}1}}}{1}
      {2}{{{\color{numb}2}}}{1}
      {3}{{{\color{numb}3}}}{1}
      {4}{{{\color{numb}4}}}{1}
      {5}{{{\color{numb}5}}}{1}
      {6}{{{\color{numb}6}}}{1}
      {7}{{{\color{numb}7}}}{1}
      {8}{{{\color{numb}8}}}{1}
      {9}{{{\color{numb}9}}}{1}
      {:}{{{\color{punct}{:}}}}{1}
      {,}{{{\color{punct}{,}}}}{1}
      {\{}{{{\color{delim}{\{}}}}{1}
      {\}}{{{\color{delim}{\}}}}}{1}
      {[}{{{\color{delim}{[}}}}{1}
      {]}{{{\color{delim}{]}}}}{1},
}

\begin{document}

\title{Distributed Attribute-Based Access Control System Using a Permissioned  Blockchain
}

\titlerunning{Hyperledger Fabric Attribute-Based Access Control System}        

\author{Sara Rouhani         \and
        Rafael Belchior      \and
        Rui S. Cruz          \and 
        Ralph Deters 
}



\institute{Sara Rouhani, Ralph Deters \at
              Department of Computer Science, University of Saskatchewan, Saskatoon, SK S7N5C9, Canada \\
              \email{sara.rouhani@usask.ca}             \\
              \email{deters@cs.usask.ca}
           \and
           Rafael Belchior, Rui S. Cruz \at
              Department of Computer Science and Engineering, Instituto Superior Técnico, Universidade de Lisboa, Portugal \\
              \email{rafael.belchior@tecnico.ulisboa.pt}             \\
              \email{rui.s.cruz@tecnico.ulisboa.pt}
}

\date{Received: date / Accepted: date}

\maketitle

\begin{abstract}
Auditing provides an essential security control in computer systems, by keeping track of all access attempts, including both legitimate and illegal access attempts. This phase can be useful to the context of audits, where eventual misbehaving parties can be held accountable.  Blockchain technology can provide trusted auditability required for access control systems.  In this paper, we propose a distributed  \ac{ABAC} system based on blockchain to provide trusted auditing of access attempts. Besides auditability, our system presents a level of transparency that both access requestors and resource owners can benefit from it.  We present a system architecture with an implementation based on Hyperledger Fabric, achieving high efficiency and low computational overhead. The proposed solution is validated through a use case of independent digital libraries. Detailed performance analysis of our implementation is presented, taking into account different consensus mechanisms and databases. The experimental evaluation shows that our presented system can process 5,000 access control requests with the send rate of 200 per second and a latency of 0.3 seconds.


\end{abstract}

\acresetall
\section{Introduction}
\label{sec:intro}
Access control systems exist to protect system resources from unauthorized accesses. Based on the system policies, security procedures within the organization, and the level of the sensitivity of the resources, the access control systems follow one of the available access control models. 

\begin{figure}[t]
    \centering
    \includegraphics[width=\columnwidth]{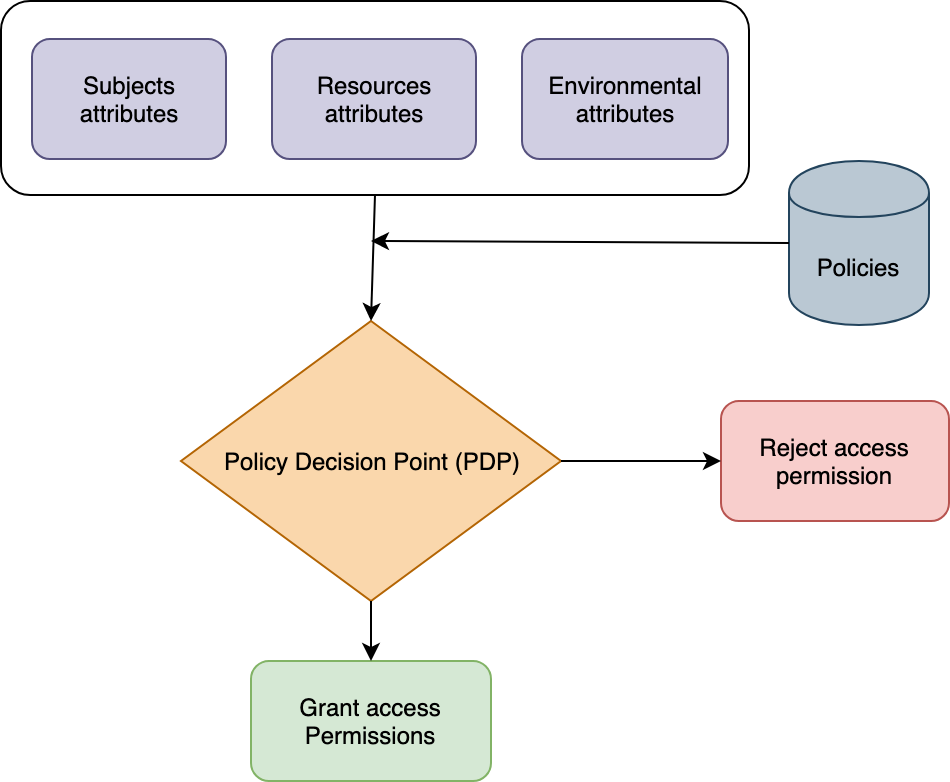}
    \caption{Attribute based access control.}
    \label{fig:abac}
\end{figure}

\ac{ABAC} \cite {hu2013guide,yuan2005attributed} is an access control model that regulates access permissions, based on the characteristics (in this context called attributes) of subjects, resources, and context (or environment).
 Access decisions are made by evaluating these attributes based on defined policies. 
\cref{fig:abac} shows the overview of the \ac{ABAC} model.

\ac{ABAC} has some advantages over other access control models as, 
\begin{inparaenum}[(a)] 
\item it can provide fine-grained and flexible access control because it allows an arbitrary number of attributes in access control decisions; 
\item the implementation of complex policies is simple and applicable; and
\item it can provide dynamic and effective access control decisions by involving environmental attributes in decision making.
\end{inparaenum}

Auditing is one of the essential controls in systems security. Auditing is the action of tracking all access attempts, including both legitimate and illegal access attempts. Keeping track of legitimate access attempts helps with non-repudiation, and keeping track of illegal access attempts helps with identifying potential threats. 
Auditability is also one of the key characteristics of blockchain by providing a trustable history of traceable transactions \cite{Belchior2019_Audits,Belchior2019}. Blockchain can exploit smart contracts to store access control policies, process access decisions,  store the result of access decisions, and accountability regarding stakeholders with different incentives. Then, at any point in the future, all access attempts toward a particular resource can be queried from the blockchain. This feature can be used as an authentic proof for non-repudiation, or it can be studied for further analysis to identify possible threats.

Besides, blockchain presents other beneficial features that are desirable for access control systems such as immutability and transparency. For example, if a  malicious system administrator changes a policy to grant or deny someone access, it will be recorded on the blockchain, and it is not possible to delete the trace of updates on policies from the blockchain. For each policy, all history of changes applied in the policy can be queried by permissioned users in permissioned blockchain. However, we prevent such a problem by configuring smart contracts so that authenticated parties must approve any change in access control policies before execution. 

In this study, we propose a complete end to end solution for implementing an \ac{ABAC} system with the focus on policy-based architecture based on Hyperledger Fabric permissioned blockchain\footnote{\url{https://www.hyperledger.org/projects/fabric}}.
Our paper contributions are summarized as follows:
\begin{itemize}
\item We propose an architecture for implementing a flexible access control system based on \ac{ABAC} and permissioned blockchain
\item We discuss our access control components including \ac{PIP}, \ac{PDP}, \ac{PAP}, which are implemented as smart contracts (\textit{or Chaincode}) 
\item We provide a specific use case of digital libraries to represent the system operation modelling. 
\item We carried out experiments, and we analyzed the performance of the presented access control application using Hyperledger Caliper  \footnote{https://www.hyperledger.org/projects/caliper} based on multiple configurations, including different databases and consensus methods.
\item Our performance analysis results also conduct a comprehensive comparison between various network configurations and pluggable components in Hyperledger Fabric modular architecture.
\end{itemize}

The remainder of this paper is organized as follows. The initial concepts of blockchain and access control systems are presented in \cref{sec:concept}, followed by  \cref{sec:related}, which reviews related studies. \cref{sec:model} explains the system model, architecture and representation of designed components.  A case study based on access to the digital libraries' resources is introduced in \cref{sec:caseStudy}. \cref{sec:evaluation} presents the evaluation results based on the represented case study and multiple configurations. Finally,  \cref{sec:conclusion and future work} concludes the paper and refers to our future work.   
\section {Background}\label{sec:concept}

\subsection{Blockchain and smart contracts}
Blockchain is a particular type of distributed ledger technology. The data is recorded on the blockchain as a group of transactions called blocks. Each block has a hash value, and it links to the previous block by referencing the hash value of the previous block in the header of the current block. Consequently, data manipulation is not possible in the blockchain, as even a slight change leads to an inconsistency between linked blocks, and can be recognized easily. In order to attach a valid block to the blockchain, a consensus mechanism is applied. There are several consensus mechanisms with a trade-off between performance and security. 

Before the development of smart contracts, blockchain applications were limited to creating cryptocurrencies and simple monetary transactions. The development of smart contracts has provided the infrastructure for creating more diverse blockchain-based applications. Smart contracts are executable logic encoded in blockchain with the ability to enforce automatically.

Blockchain networks can be divided into two main categories: public and permissioned.

Public blockchains are open to the world, and every user can join the blockchain with an anonymous identity, submit a transaction, and participate in consensus. Permissioned blockchains include an additional membership layer, so only authenticated users can join the blockchain and interact with different components. 

Today many blockchain platforms exist, and they are geared toward implementing smart contracts and decentralized applications. They are different in different aspects, such as the type of the network(public or permissioned), built-in cryptocurrency, transaction workflow, performance, privacy, cost and, most importantly, maturity. Some blockchain platforms such as Ethereum, Hyperledger Fabric, Corda have mature tools, while others have very little support for their users and developers.

Hyperledger Fabric~\cite{Androulaki2018} is a famous implementation of a permissioned blockchain, hosted by the Linux Foundation. Hyperledger Fabric has a modular structure that allows component pluggability, such as consensus, membership, and database. The membership layer can authenticate users and grant access to users based on their access level and system policy. Hyperledger Fabric has integrated the \ac{ABAC} mechanism, so it is possible to build permission groups for access control by checking members' attributes. However, access control parameters and permission groups have to be predefined. It is not suitable for applications that require dynamic and flexible access control.

Our presented system provides a solution for off-chain parties that look for a flexible and distributed access control service compatible with their authentication service. Our provided solution can be easily integrated with any off-chain system, while access control functionality is achieved through blockchain and smart contracts.

Smart contracts correspond to logic encoded in the blockchain that can be programmed and deployed as an automation program. Accordingly, they can create complex transactions and enforce their conditions automatically~\cite{Rouhani2019}.

Chaincode is a term introduced by Hyperledger Fabric for smart contracts. Chaincode may consist of multiple smart contracts or include only one smart contract. We use the chaincode and smart contract concepts interchangeably.

Before the development of smart contracts, blockchain applications were limited in creating cryptocurrencies and simple monetary transactions. The development of smart contracts provided the infrastructure for creating more diverse blockchain-based applications, such as healthcare \cite{kuo2017blockchain,Azaria2016,Dagher2018}, \ac{IoT} \cite{Khan2018,Novo2018}, resource sharing \cite{Zhu2018,Wang2018}, and business process management \cite{weber2016untrusted,lopez2017caterpillar,sarabpm}. In our previous paper \cite{Rouhani2019a}, we discussed that although the applications of these systems are different, their primary purpose is similar as they aim to control access over particular data. The domain of the data is their main difference; for example, it could be patient healthcare data or data generated by \ac{IoT} devices.

\subsection{Access control models and \ac{ABAC}}
Access control refers to any action to prevent data and resources from unauthorized access, disclosure or modification. In traditional databases, an authorization defined by a triplet $<o, s, p>$ and defines that subject $s$ is authorized to execute privilege $p$ on object $o$ \cite{BERTINO19941}.

Conventional access control models follow such definition, adding into consideration the context in which an access control request is performed.

\begin{inparaenum}[(1)]
\item { \ac{DAC} \cite{biba} or authorization-based}, 
\item {\ac{MAC} \cite{Bell1976} }, and later 
\item {\ac{RBAC} \cite{Ferraiolo:01} }
\end{inparaenum} are three initial access control models \cite{samarati}. 

\ac{DAC} restricts access permissions based on the subjects' identity, and the resource owner defines policy rules. 

In \ac{MAC}, the system defines access policies through the security labels. This model usually is used for controlling access over sensitive and confidential data. 

In \ac{RBAC}, there are predefined roles in the system and users have different access levels depending on their roles. 

\ac{ABAC} is logical access control that comprises access control lists, role-based access control, and its own method for providing access based on the evaluation of attributes \cite {hu2013guide}. \ac{ABAC} controls access to the system resources by evaluating policies (system rules) against entities' attributes, including subject, object, and environmental attributes. Attributes are characteristics of the subjects (users) and protected objects (resources). The environment conditions as the environment's attributes can also be taken into account for \ac{ABAC} decision making. 

\ac{ABAC} is a flexible and fine-grained mechanism that is also capable of enforcing the other three methods. Distributed systems also adopted \ac{ABAC} as they require federation and autonomy control among coordinated systems, and \ac{ABAC} enables granular and meta attribute capabilities that support privilege delegation in a distributed application~\cite{Hu2018}.
Likewise, blockchain-based applications mostly adopt \ac{ABAC} \cite{Rouhani2019a}. 

XACML \cite{Anderson2006} introduces a policy-based architecture for the specification and enforcement of access control policies. The architecture comprises the following components. 
\begin{itemize}
    \item \textit{Client}: the device that requests access to a resource, possibly on behalf of a user. 
    \item \textit{\acf{PEP}}: the network device on which access decisions are carried out. PEP serves as the gatekeeper to the intended resource.
    \item \textit{\acf{PIP}}: the repository that holds information (attributes) about the client and provides this information to the PDP. 
    \item \textit{\acf{PDP}}: the component that decides to allow or deny the client access to the resource. 
    \item \textit{\acf{PAP}}: the component that is responsible for managing access control policies. 
    \item \textit{Accounting or Auditing}: The component that is responsible for tracking access attempts.
\end{itemize}

\begin{figure}[ht]
    \centering
    \includegraphics[width=1\columnwidth]{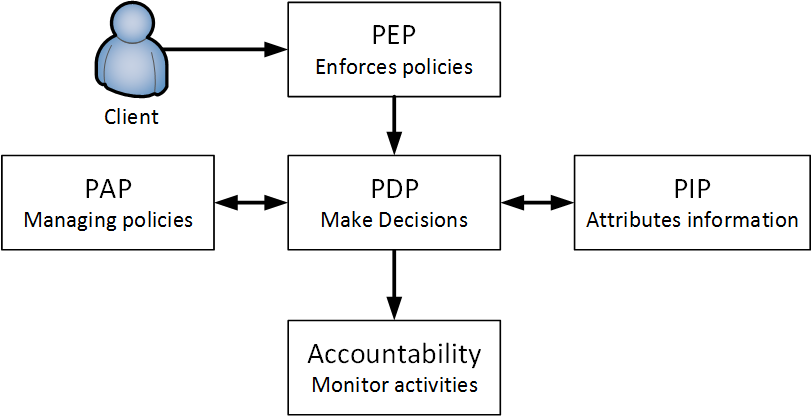}
    \caption{ABAC logical components based on Policy based architecture \cite{Anderson2006}\label{fig:Policy based architecture}}
\end{figure}

 \Cref{fig:Policy based architecture} illustrates how these components interact with the client and each other. A client requests access permission, and \acf{PEP} forwards the request to \acf{PDP}.  \acf{PDP} queries the related policy and attributes from \acf{PAP} and \acf{PIP}. After receiving the required information, \acf{PDP} assesses the access decision and sends the result to \acf{PEP} for enforcing the access decision.

\section{Related Work}\label{sec:related}
Many studies on blockchain technology focus on presenting an access control system either in the context of specific applications, such as healthcare \cite{zhang2018blockchain,saraMedichain,Azaria2016,xia2017medshare,Dagher2018,rajput2019eacms}, \ac{IoT} \cite{novo2018blockchain,DBLP:journals/corr/abs-1802-04410,dukkipati2018decentralized,Pinno2018,ouaddah2016fairaccess,Ding2019,ma2019privacy}, and cloud federation \cite {ferdous2017decentralised,alansari2017distributed} or they introduce a general access control system, which can be employed for different applications. In our previous study \cite{Rouhani2019a}, we have investigated the state of the art of access control systems based on blockchain. In this section, we overview similar studies, which present an attribute-based access control based on blockchain. As illustrated in table \cref{table:relatedworks}, many studies use the attributed-based method for their access control system because of the granularity, flexibility, and dynamic features that \ac{ABAC} provides. 

\begin{table*}[h]
\label{table:relatedworks}
\centering
\caption{A summary of blockchain-based access control applications.}
\resizebox{\textwidth}{!}{%
\begin{tabular}{|c|c|c|c|c|}
\hline 
Research paper  & Domain    & Access control method                                                                      & Privacy Support & Scalability \\ 
\hline

Jemel and Serhrouchni \cite{jemel2017decentralized} & Data sharing              & \ac{ABAC} + Attribute-based Encryption  & {\checkmark}   & {x}           \\ 
\hline

Wang \textit{et al.} \cite{wang2018blockchain} & Data sharing                   & \ac{ABAC} + Attribute-based Encryption  & {\checkmark}   & {x}           \\ 
\hline

Zhu \textit{et al.} \cite{zhu2018tbac}	& Resource sharing               	    & \ac{ABAC}                               & {\checkmark}   & {x}    	   \\ 
\hline

Hu \textit{et al.} \cite{hu2018reputation}       & Knowledge sharing            & Fine-grained                            & {x}            & {x}           \\ 
\hline

Ferdous \textit{et al.} \cite{ferdous2017decentralised}       & Cloud federation & -                                      & {x}            & {x}            \\ 
\hline

Alansari \textit{et al.}  \cite{alansari2017distributed}     & Cloud federation & \ac{ABAC}                               & {\checkmark}   & {x}            \\ 
\hline

Zhang and Posland  \cite{zhang2018blockchain}   & Health care                   & \ac{ABAC}                               & {\checkmark}   & {x}             \\ 
\hline

Rouhani \textit{et al.} \cite{saraMedichain}   & Health care                    & Role-based                              & {\checkmark}   & {\checkmark}   \\ 
\hline

Guo \textit{et al.} \cite{guo2019access} & Health care               & \ac{ABAC}                               & {x}            & {\checkmark}  \\ 
\hline
Asaph \textit{et al.}  \cite{Azaria2016}        & Health care                   & -                                       & {\checkmark}   &  {\checkmark}  \\ 
\hline

Xia \textit{et al.}   \cite{xia2017medshare}      & Health care                 & -                                       & {\checkmark}   & {x}            \\ 
\hline

Dagher \textit{et al.} \cite{Dagher2018}		& Health care         	       & Role-based                               & {\checkmark}   & {\checkmark}	 \\ 
\hline

Rajput \textit{et al.} \cite {rajput2019eacms}  & Health care                  & Role-based                               & {\checkmark}   & {\checkmark}     \\ 
\hline

Zyskind \textit{et al.} \cite{Zyskind}          & Mobile applications          &  Policy-based                            & {\checkmark}   & {\checkmark}     \\ 
\hline

Novo       \cite{novo2018blockchain}           & IoT                           & -                                        & {x}            & {\checkmark}     \\ 
\hline

Outchakoucht \textit{et al.} \cite{outchakoucht2017dynamic} & IoT              & Policy-based                             & {x}            & {x}               \\ 
\hline

Zhang \textit{et al.} \cite{DBLP:journals/corr/abs-1802-04410} & IoT           & Policy-based and dynamic access control  & {x}            & {x}               \\ 
\hline 

Dukkipati \textit{et al.} \cite{dukkipati2018decentralized}		& IoT	       & \ac{ABAC}                               & {\checkmark}    & {x}	          \\ 
\hline

Pinno \textit{et al.} \cite{pinnocontrolchain}	& IoT 	                       & \ac{ABAC}                               &  {\checkmark}   &  {\checkmark}	\\ 
\hline

Ouaddah \textit{et al.}   \cite{ouaddah2016fairaccess}     & IoT               & -                                       &  {\checkmark}  &  {\checkmark}    \\ 
\hline

Ding \textit{et al.} \cite{Ding2019} & IoT                                    & \ac{ABAC}                                & {x}            & {\checkmark}   \\ 
\hline

Ma \textit{et al.} \cite{ma2019privacy} & IoT                                 & Generic                                  & {\checkmark}   & {\checkmark}     \\ 
\hline

Rouhani \textit{et al.}   \cite{sara_physical_ac_mang} & Physical access control & Role-based                            & {x}          & {\checkmark}       \\ 
\hline

Es-Samaali \textit{et al.}   \cite{es2017blockchain} & Big data management     & \ac{ABAC}                               &  {\checkmark}     &  {\checkmark}  \\ 
\hline

Xh \textit{et al.} \cite{xu2019exploration} & Space situation awareness       & Capability-Based                         & {x}               & {\checkmark} \\ 
\hline
Lyu \textit{et al.} \cite{lyu2020sbac} & Information centric networking & matching-based & {\checkmark} & {x}    \\ \hline
Paillisse \textit{et al.} \cite {paillisse2019distributed} & Multi-administrative domain & -                             & {\checkmark}       & {x} \\  
\hline
Laurent \textit{et al.} \cite{secrypt18bbacs}      &  General access control    &  Access Control List                    & {\checkmark}   & {x}           \\ 
\hline

Maesa \textit{et al.} \cite{DiFrancescoMaesa2019a} & General access control     &  \ac{ABAC}                              & {x}            & {\checkmark}  \\ 
\hline

Guo \textit{et al.} \cite{guo2019multi} & General access control                & \ac{ABAC}                               & {\checkmark}   & {x}           \\ 
\hline

Lee \textit{et al.} \cite{lee2019blockchain} & General access control           & Role-based                              & {\checkmark}   & {x}           \\ 
\hline
\end{tabular}}
\end{table*}

Guo \textit{et al.} in~\cite{guo2019access} introduce a hybrid architecture for access control over \ac {EHR} data using blockchain and edge nodes. The blockchain acts as a tamper-proof validation component to verify identities and access control policies. The edge nodes store the \ac {EHR} data off-chain and enforce the access controls. 
The smart contracts include the address of \ac {EHR} data on the edge nodes by using one-time self-destructing URLs \footnote{https://1ty.me/}.
Based on performance results against unauthorized retrieval for the average transaction processing time was 40 ms, and the average response time was 30 ms. Also, the test result based on a high number of patients does not affect the response time and indicate the scalability of their solution. 

Zyskind \textit{et al.} in~\cite{Zyskind} conceptualizes the blockchain technology as an access control moderator, complemented by an off-blockchain storage solution. Blockchain clients representing users that provide their data to a service provider are the owners of their data. Based on that premise, this solution is meant to empower users, so they have the information about which data is collected about them by third parties and how their data is used. For achieving that goal, each data owner can issue transactions, used to change the set of permissions granted to a service or entity. Each transaction is recorded on the blockchain, allowing for auditability and traceability. 

Zhang \textit{et al.} in~\cite{DBLP:journals/corr/abs-1802-04410} propose a solution directed to \ac{IoT} blockchain-based access control. The authors introduce the concepts of \textit{Judge Contract} (JC), \textit{Register Contract} (RC) and \textit{Access Control Contracts} (ACC). Access control contracts store access control policies for a subject-object pair. In this system, both JC and RC are essential pieces in regards to achieving a distributed and reliable access control. The JC receives misbehaviour reports and applies penalties according to them. The RC stores the misbehaviour information from the JC and manages it through the judging method. Moreover, it stores information such as name, subject, object, and smart contract for access control.

Zhu et al. in~\cite{zhu2018tbac} propose a \ac{TBAC} system that integrates \ac{ABAC} model into the bitcoin blockchain. There are four implemented transactions, including subject registration, object escrowing and publication, access request and grant. They also present a cryptosystem associated with their system as an additional security layer. The system is evaluated in terms of security, but the performance and scalability of the system are not examined. 

In the federated cloud services, access control enforcement is still vulnerable to privacy violations. 

\sloppy Alansari \textit{et al.} in~\cite{alansari2017distributed} present an attribute-based access control system based on Pedersen commitment scheme~\cite{pedersen1991non} and blockchain. The system is designed to keep the users' attributes private from the federated organization. Users' identity attributes and access control policies are stored on the blockchain to guarantee the integrity of them. They also employed Trusted hardware technology to guarantee the integrity of the policy enforcement process. 

Zhang and Posland in~\cite{zhang2018blockchain} propose an architecture for granular access authorization that supports flexible queries, which provides secure authorization at different levels of granularity. The designed architecture offers a capable infrastructure without requiring the public key infrastructure (PKI), so it decreases the computation time needed and suitable for the devices with limited resources in EMR systems. As a result, their system can efficiently respond to a requester without exposing unauthorized private data.

Maesa \textit{et al.} in~\cite{DiFrancescoMaesa2019a} implemented an access control service on top of Ethereum \footnote{\url{www.ethereum.org}}. Blockchain is used to store smart contracts that represent access control policies represented in \ac{XACML}~\cite{Anderson2006} and to perform the decision process. Such smart contracts are called \textit{\acp{SP}}. Thus, \acp{SP} are responsible for the policy evaluation process, embedding a \ac{PDP} for a specific access control policy.  Each time, an access request needs to be evaluated to make an access decision, and the blockchain executes it in a distributed way. The decision is made based on information concerning the users. For this purpose, the concept of \ac{AM} is introduced. \acp{AM} are the components that manage the attributes of the entities involved in the process, such as subjects, resources, and environmental context. \acp{AM} can update and retrieve their values and are created by an entity, the \ac{AP}. Implementation based on Ethereum blockchain is costly since, for every operation, a fee called gas must be paid. Although for public blockchain systems, it is generally unavoidable, for permissioned implementation, this is an unnecessary additional cost imposed on the system.

The privacy and security of the data generated by \ac{IoT} devices are the major concern of the IoT system due to the extensive scale and distributed nature of \ac{IoT} networks. In order to protect users' privacy, many studies have considered blockchain to provide secure access control to the \ac{IoT} data. References~\cite{dukkipati2018decentralized,pinnocontrolchain,Ding2019} presented an \ac{ABAC} for \ac{IoT} systems. Dukkipati \textit{et al.} solutions \cite{dukkipati2018decentralized} store system policies off-chain while \cite{pinnocontrolchain} stores the polices on the Ethereum platform and is less vulnerable to security breaches, but more costly. 
Ding \textit{et al.} in~\cite{Ding2019} focus on simplifying the access control protocol to make it lightweight and suitable for \ac{IoT} devices with limited computing capability and energy resources.

\begin{figure}[ht]
    \centering
    \includegraphics[width=1\columnwidth]{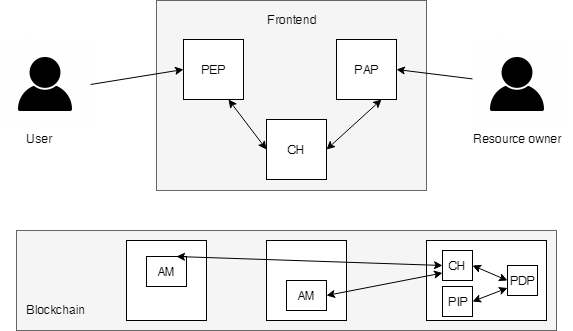}
    \caption{Blockchain-based access control service architecture~\cite{maesa2017blockchain}.
    \label{fig:bac-xacml}}
\end{figure}

Zhang and Posland in~\cite{Zhang2018} present an architecture for a blockchain-based \ac{EMR} access with granularity control that supports flexible data queries. The user layer first sends a query that could have different levels of granularity, such as block query, attribute query, or mixed query. The agent layer aggregates the query data and authorizes the user, who has access permission for that query. If it is a valid query, it passes the query to the storage layer. The storage layer returns the data to the agent layer. The agent layer encrypts the query and sends it to the user layer. Lastly, the query is decrypted, using the provided access keys. The represented architecture provides a flexible infrastructure to achieve granular access control without requiring \ac{PKI}, so it leads to a decrease in computation time proceeding. 

As we reviewed, many studies have investigated blockchain as a back-end infrastructure for the distributed access control system. However, most of the prior works in this area are domain-specific. It means their access control solutions are designed for a particular domain, such as healthcare data or IoT data. Besides, most of these studies lack the details of implementation and performance analysis. As a result, it remains unclear if a blockchain can be the basis for access management at large scale.

\section{System Model and Architecture} \label{sec:model} Centralized access control systems suffer from various problems such as:
\begin{inparaenum}[(a)]
\item the risk of privacy leakage, and 
\item the risk of a single point of failure;
\item interoperability issues;
\item unreliability of the access control system, and
\item the presence of third parties.
\end{inparaenum}

An access control system can utilize the blockchain technology to address these problems.
The decentralized nature of the blockchain resolves the problem of a single point of failure. Cryptographic methods ensure the reliability of the ledger.
Consensus mechanisms ensure that the state of the ledger is valid, and it is the same for every participant. Smart contracts allow monitoring and enforcement of sophisticated access control decisions. Also, with automatic enforcement, they can address privacy issues.

This solution empowers both resource owners and subjects (typically access requesters). Details of each granted or revoked access permission can be queried from the ledger, including the policies that have been applied, the attribute values and the time of access request.  
 In practice, under no circumstances, resource owners do not deny access to a resource by a rightful requester. On the other hand, the resource owners leverage provided audit trails, while being assured that no user had subverted the system.

\begin{figure}[h]
    \centering
    \includegraphics[width=1\columnwidth]{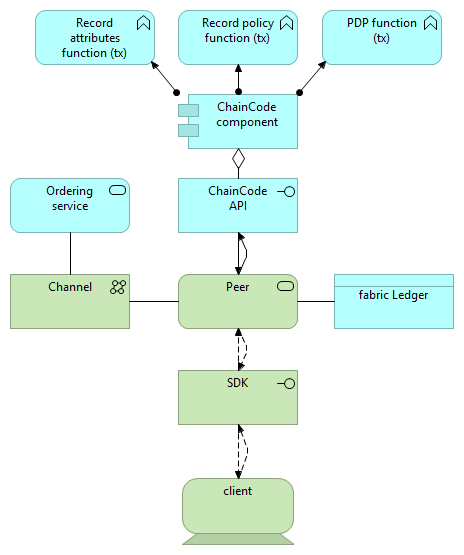}
    \caption{High level system architecture using the Archimate modeling language \cite{group2016archimate}.}
    \label{fig:model}
\end{figure}

To provide a solution to these problems, we used the Hyperledger Fabric blockchain. In the blockchain, there have to be at least two endorsing nodes belonging to different organizations. These nodes are responsible for executing \acp{SP}. Clients would be the systems that use this system, as depicted in~\cref{fig:model}. 

The workflow of the users remains unaltered. The only difference is that the authorization requests are now mediated through one or more nodes representing the given system, as the evaluation of access control policies could be not trusted for the subject of the request, who instead requests invalid access to resources. 

\begin{figure*}[ht]
\centering
\includegraphics[width=1\textwidth]{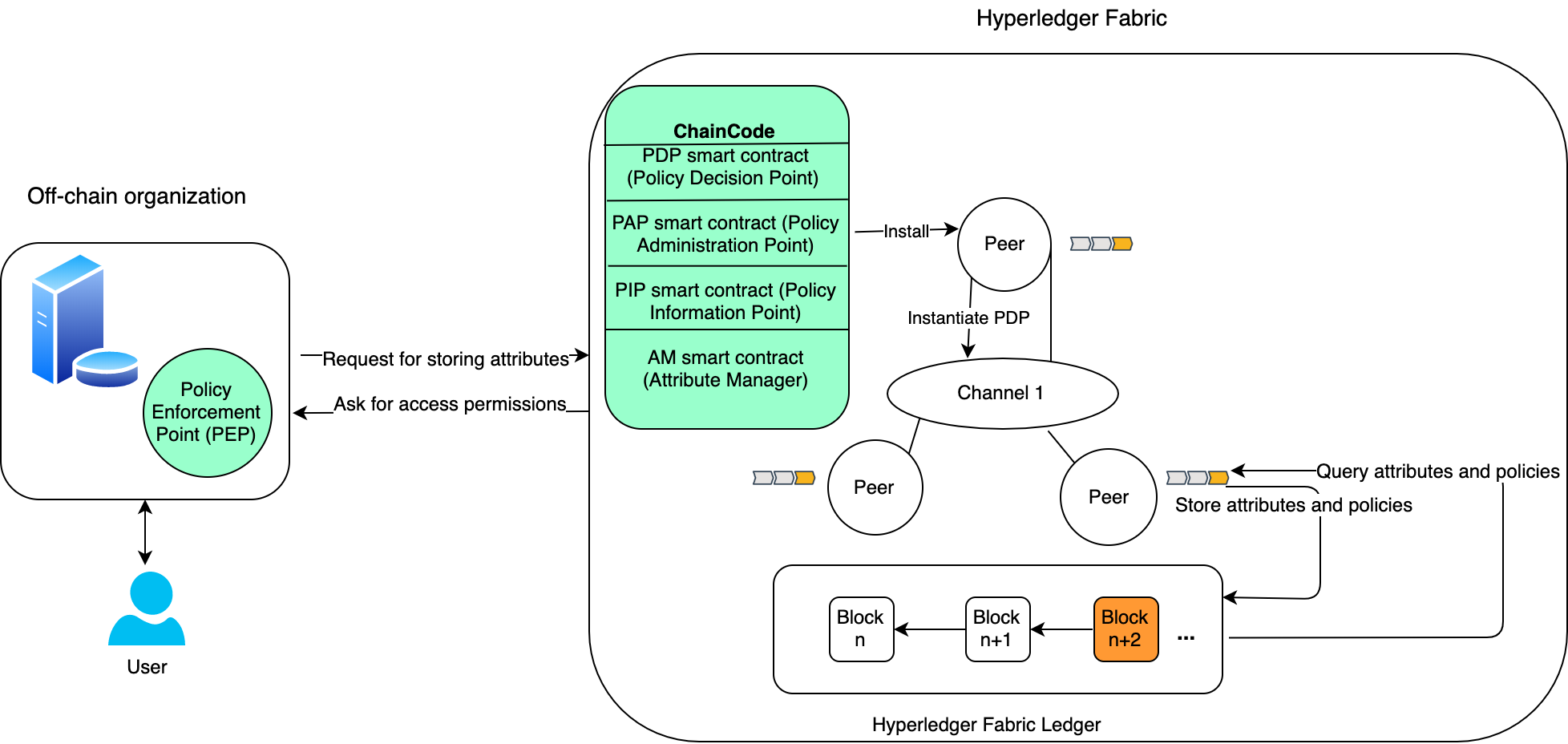}
\caption{Blockchain Access Control System Architecture.\label{fig:architecture}}
\end{figure*}

 In order to protect the privacy of users' data, Hyperledger Fabric provides a private data feature to protect sensitive users' data. We have used this feature to represent the attributes that are required for access permissions based on the organizations' defined policies. The private data is hashed, and then it will be endorsed and ordered like other data. Finally, the chaincode writes hashed data on the ledgers of every peer. However, only organizations that require these private attributes to give access permission have access to them. Using Zero-knowledge proofs \ac{ZKP}\cite{Goldreich94definitionsand} is another alternative that can be applied for highly privacy-preserving case studies that require to protect all users' attributes from all access providers and data owners. However, \ac{ZKP} requires additional time and computational resources compared to our solution based on Hyperledger Fabric private data feature.

 \label{sec:architecture}
In our solution, as depicted in Figure \ref{fig:architecture}, the blockchain acts as a mediator between the entity that requests access to a specific resource and the entity that manages that resource. The system includes two main components. The first component is an off-chain system that relies on permissioned blockchain to store its access control attributes on it and query access permissions from it. The second component is a permissioned blockchain that manages different access control components through smart contracts and stores the data on a tamper-proof ledger.

\begin{figure}
    \centering
    \includegraphics[width=1\columnwidth]{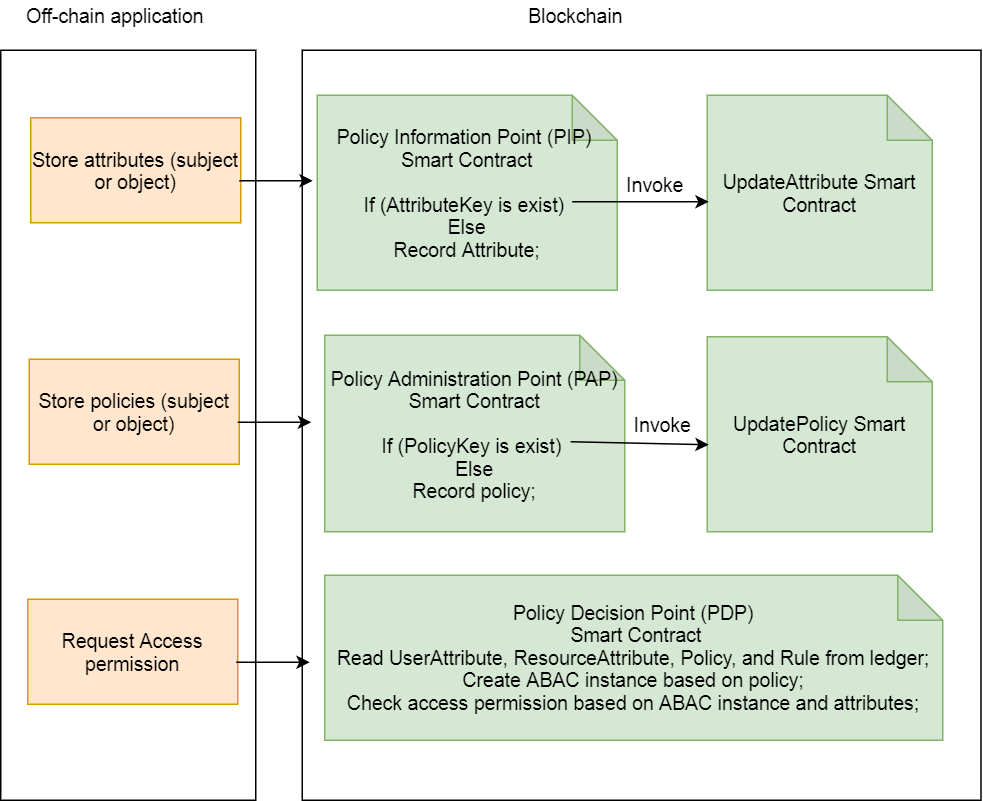}
    \caption{Smart contracts architecture}
    \label{fig:smartContracts}
\end{figure}

The three main smart contracts are \ac{PIP} contract, \ac{PAP} smart contract and the \ac{PDP} contract. Subject (users) and resource (objects) attributes are stored in a \ac{JSON} data format through the \ac{PIP} contract. The \ac{PIP} is also responsible for checking write conflicts and updating attributes. Policies are also recorded in the system as \ac{JSON} data format, and \ac{PAP} contract is responsible for managing policies and updating policies. The system could be implemented to work with multiple \ac{PAP}run by different organizations. Even if there is only one \ac{PAP}, the transparency offered by this solution distributes the trust and the responsibility of these access policies. \ac{PDP} contract evaluates policies to make an access decision. \Cref{fig:smartContracts} shows the architecture of the implemented smart contracts.

 After smart contracts evaluate \acp{SP} against their respective attributes, the \ac{PDP} returns its decision to the \ac{PEP}. This process allows a decoupling between users and the blockchain administration (as users do not need to have a node on the blockchain, which is desirable).

 Our solution not only logs all access requests in a very secure way but also provides a framework to control all access controls concerning the participants in the network. Nodes from the private network can access the blockchain, check transactions' history, and audit the history of access requests and results. Automatic auditing techniques can be developed by analyzing the history of access request transactions. A fine-grained access control solution is provided that enforces access validation through blockchain-based service providers.
\section{Case Study} \label{sec:caseStudy}
A digital library is a collection of documents in an organized electronic form that allows users to access them online. A highly dynamic user population and the numerous collection of resources in digital libraries require a fine-grained and dynamic access control method such as ABAC \cite{adam2002content}. It requires that access policies specified based on users' attributes and characteristics rather than users' roles in the system. 

In this section, we have selected a case study for access control in digital libraries to illustrate and explain the application of our \ac{ABAC} system.

For every subject, we store it with a subject ID (SID), and the set of its attributes and values, and the same for Objects attributes. Attribute ID is a required field to store attributes in the Hypeledger Fabric database and later retrieve the respective attribute for permission decision. Both Hyperledger Fabric supporting databases (CouchDB and Level DB) are key-value stores, and the ID field is used as keys.

$S_nA$ is the set of attributes associated with the $subject_n$ and  $S_nID$ defined as a key for storing the correlated attributes. Similarly, $O_nA$ is the set off attributes associated with the $object_n$ and  $O_nID$ defined as a key for storing the correlated attributes. 

$P_nSA$  and $P_nOA$ are the sets of subjects and Objects determinative attributes in the policy of n. Same as attributes, Policy ID ($P_nID$) is a required field to store attributes. For every policy, we store the determinative attributes ($P_nSA, P_nOA$) along with the policy's rules.

$S_nA = \{S_nID, \{\{S_nA_1, value \}, ... , \{S_nA_n, value\}\}\}$

$O_nA = \{O_nID, \{\{O_nA_1, value \}, ... , \{O_nA_n, value\}\}\}$

Defining $SA$ as the set of all subjects attributes and $OA$ as the set of all objects attributes, we have:

$SA = (S_1A \cup S_2A \cup ...S_nA)$

$OA = (O_1A \cup O_2A \cup ...O_nA)$

$P_nSA \subseteq	 SA$

$P_nOA \subseteq OA$

$P_nSA, P_nOA \in P$

$P_n = \{P_nID, P_nSA, P_nOA, rules\}$

Resulting, we have the following attributes for the subject $S_1A$ and object $O_1A$:

$S_1A = \{``s001", \{``status", true \}, 
\{``expiration",
``2020-05-12"\},\{``libraryGroup",  12 \}\}$

$O_1A = \{``r001", \{``libraryGroup",  12 \}\}$

The policy $P_1$ with the id ``policy01" is as following: 

$P_1= \{``policy01", S_1A, O_1A, \{``status==true" \land ``expiration">``1Day" \land ``user.libraryGroup"== ``resource.libraryGroup"\}\} $

\subsection{JSON Data Format}
In our implemented solution, the access control data (attributes and policies) is followed by the \ac{JSON} format. Using JSON, as a widespread data format, can be used by a broad range of applications. An example of a sample policy, including subject (user) attributes, and object (resource) attributes, is illustrated in the following \ac{JSON} code snippet. Based on the presented policy, the subject and the resource attributes, our subject has valid access permission to the resource, as the subject has valid \ac{ID}, active status, non-expired membership and the subject library group matches with the resource library group. If one of the subject's attributes does not pass the policy rules, the access permission will be denied. 

\begin{lstlisting}[language=json,numbers=none, basicstyle=\footnotesize, title={Example: Definition of a sample Access Control Policy, Subject and Resource followed by the JSON format}, captionpos=t]
policy = {
``policyID'': ``policy01'',
attributes: {
``user'': {
    ``status'': ``Active'',
    ``expiration'': ``Date of expiration'',
    ``libraryGroup'': ``Group ID''
         },
``resource'': {
    ``libraryGroup'': ``Group ID''
    }
                },
``rules'': {
    ``user.status'': {
    ``comparison_type'': ``boolean'',
    ``comparison'': ``boolAnd'',
    ``value'': true
                },
    ``user.expiration'': {
    ``comparison_type'': ``datetime'',
    ``comparison'': ``isMoreRecentThan'',
    ``value'': ``1DAY''
                },
    ``user.libraryGroup'': {
    ``comparison_target'': ``libraryGroup'',
    ``comparison_type'' ``numeric'',
    ``comparison'': ``isStrictlyEqual'',
    ``field'': ``resource.libraryGroup''
                }
            }
        }
    }
}

subject = {
subjectID: ``s001'',
attributes:{
       ``status'': true,
       ``expiration'': ``2020-05-12'',
       ``libraryGroup'': 12
}
}
resource = {
resourceID: ``r001'',
Attributes:{
    ``libraryGroup'': 12
}
}
\end{lstlisting}

\section{System evaluation} \label{sec:evaluation}
In this section, we evaluate the proposed system performance. We used Hyperledger Caliper to measure the performance of our system based on our own written benchmark and various configurations. 
\subsection{Environment configuration, performance parameters, and assumptions}
We evaluate every component of our system against two different databases, Couchdb and Goleveldb and two orderer services, Raft and Kafka. We also present the results of the evaluation based on the Solo orderer to illustrate the effect of other parameters separated from the effect of the involved consensus method.

Raft is a \ac{CFT} ordering service based on the implementation of Raft protocol. Raft follows the ``leader and follower'' model. The leader makes decisions, and the followers follow the leader. The peer that represents the leader is changed frequently. Every follower has the chance to be a candidate to become the leader of the next round.  

Similar to Raft, Kafka is a \ac{CFT} ordering service, which follows the ``leader and follower'' model as well. However, Kafka uses ZooKeeper \footnote{https://zookeeper.apache.org/} to manage clusters. Zookeeper keeps track of the status of the Kafka cluster nodes and partitions.  

Solo is a single ordering node, and it is not fault-tolerant. It is meant to be used for testing purposes.  

We used the Google Cloud Platform to run a \ac{VM} instance and test our application and collect performance analysis data. The machine type is n2-highcpu-8, which includes eight virtual CPUs and 8 GB of memory. All the tests are run on the same virtual machine, as Caliper emulates workload distribution between several clients.

The default number of blockchain clients is 10 (each client emulated by a different NodeJS\footnote{https://nodejs.org/en/} process), the default number of transactions is 5,000, and the default transaction type is policy decision transaction, which queries the related data from ledger based on access request and determines the result of the access request. The default database for Raft and Kafka is GoLevelDB. The default number of organizations is two, and the default number of peers is one. 

\subsection{Performance evaluation results}
\begin{figure}[ht]
\centering
\includegraphics[width=1\columnwidth]{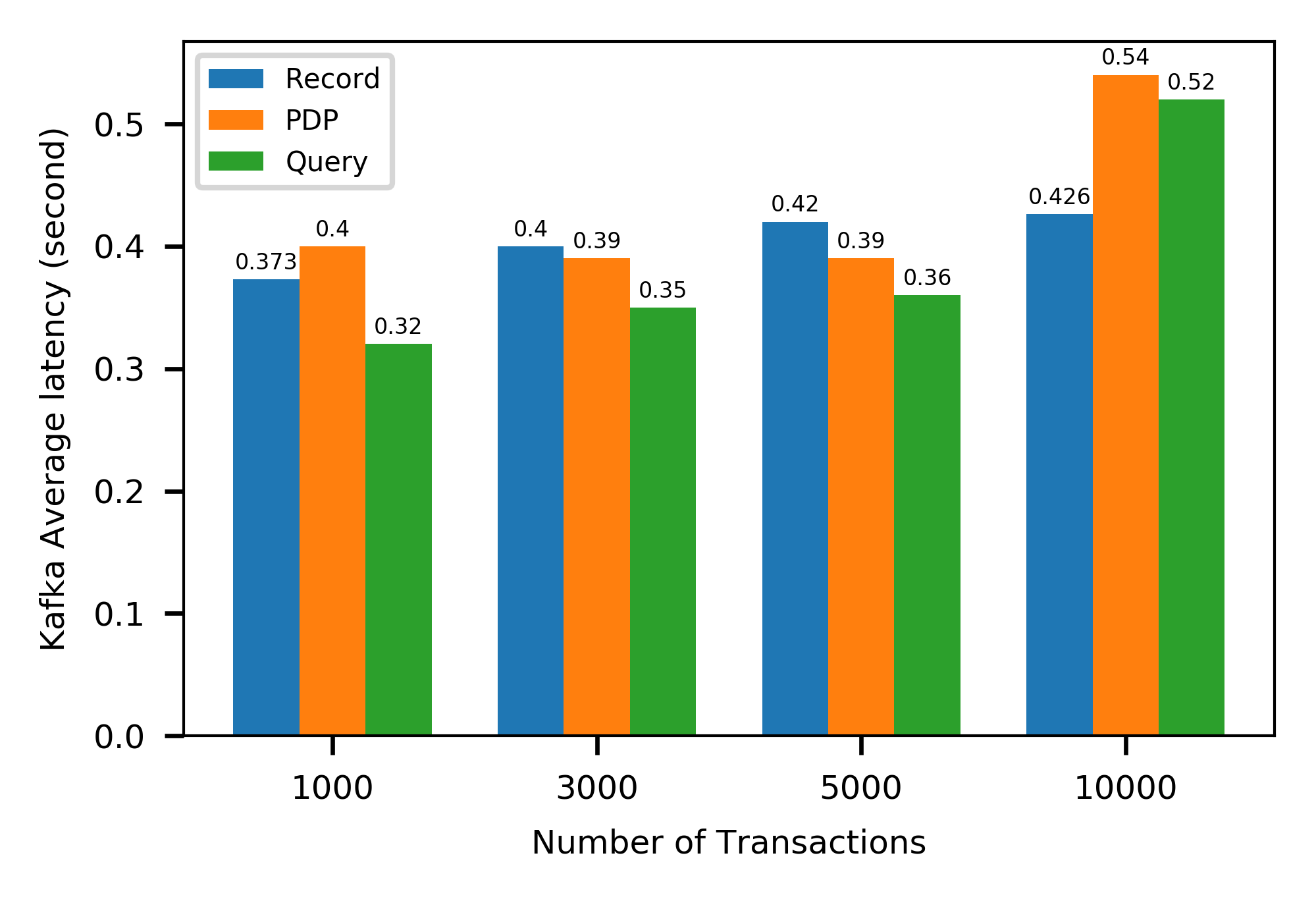}
\caption{Average latency of Kafka as a function of the type of transactions.
\label{fig:kafka-diffTX}}
\end{figure}

\Cref{fig:kafka-diffTX} shows the average latency (in seconds) for three different transactions, Record attributes, \ac{PDP}, and query data from the ledger based on Kafka orderer. In every round of the test, we configured the test with a different number of transactions. Although the average latency increases with the number of transactions, the increase is not sharp, and it increases very slowly. As the graph illustrates, the system can process 10,000 access decisions with an average latency of 0.54 seconds. 

\begin{figure}
    \centering
    \includegraphics[width=1\columnwidth]{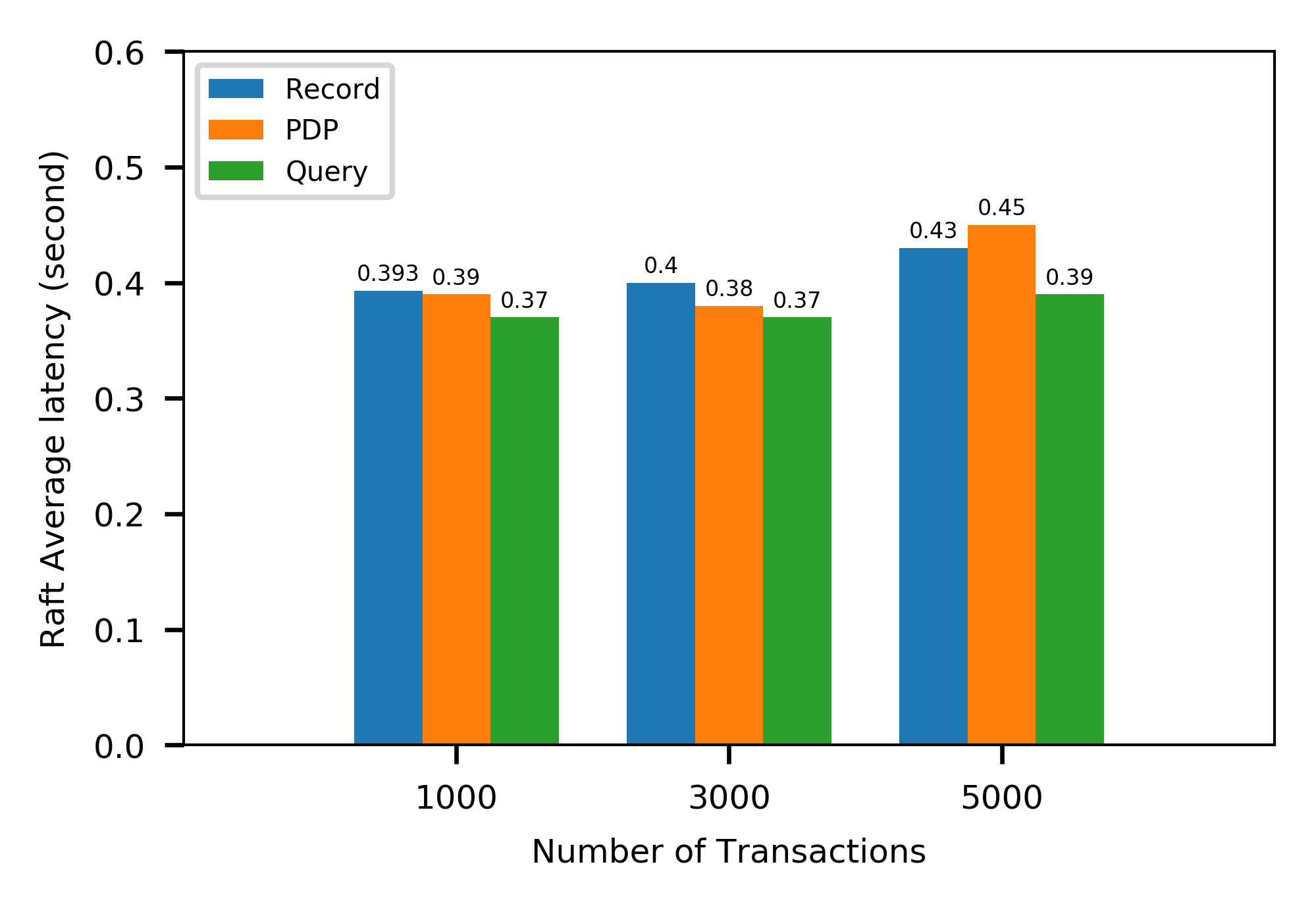}
    \caption{Average latency of Raft under different transactions.}
    \label{fig:raft-diffTX}
\end{figure}

\Cref{fig:raft-diffTX} shows the average latency (in seconds) for the same three different transactions, based on the Raft orderer. The test result is based on a different number of transactions. Similarly, the average latency increases with the increase in the number of transactions. The test run has failed for the Raft orderer with 10,000 transactions due to \ac{VM} Memory limitation. The resource consumption result indicates that Raft memory consumption increases 4.33 times, in comparison with Kafka. This fact explains why the test failed in the middle of executing 10,000 transactions with the Raft orderer. 

For both Raft and Kafka orderers, increasing the number of transactions increases the average latency for the record attributes transaction. Policy decision transaction has the minimum average latency for both Kafka and Raft, based on 3,000 transactions. For Record attribute and query data transactions, the average latency in Raft is slightly higher than Kafka. For the policy decision transaction, the average latency in Raft for 1,000 and 3,000 transactions is slightly lower than Kafka, but for 5,000 transactions, the result is the opposite.

\begin{figure}
    \centering
    \includegraphics[width=1\columnwidth]{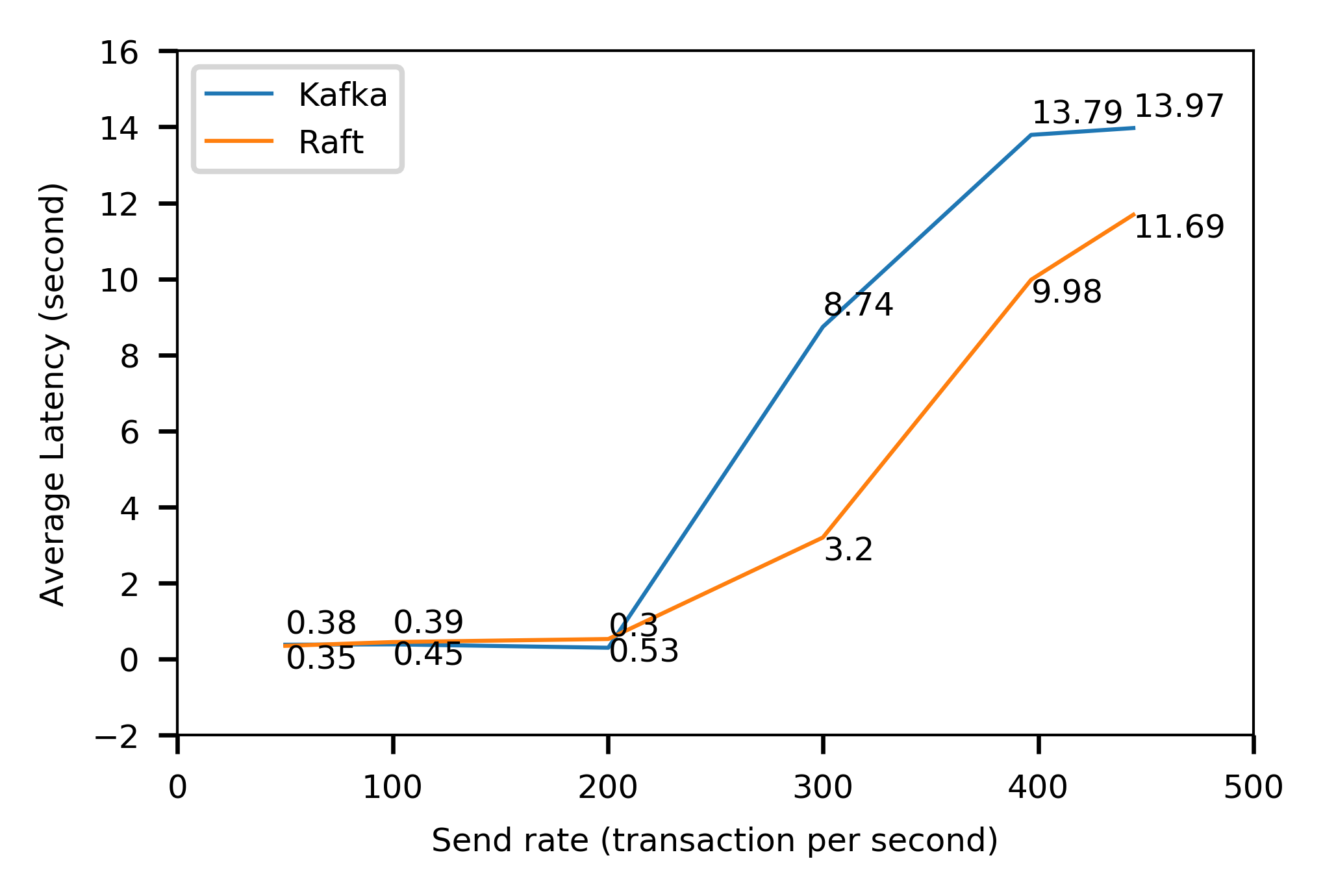}
    \caption{Average latency of Raft and Kafka based on different transactions.}
    \label{fig:raft-kafka-latency}
\end{figure}


\begin{figure}
    \centering
    \includegraphics[width=1\columnwidth]{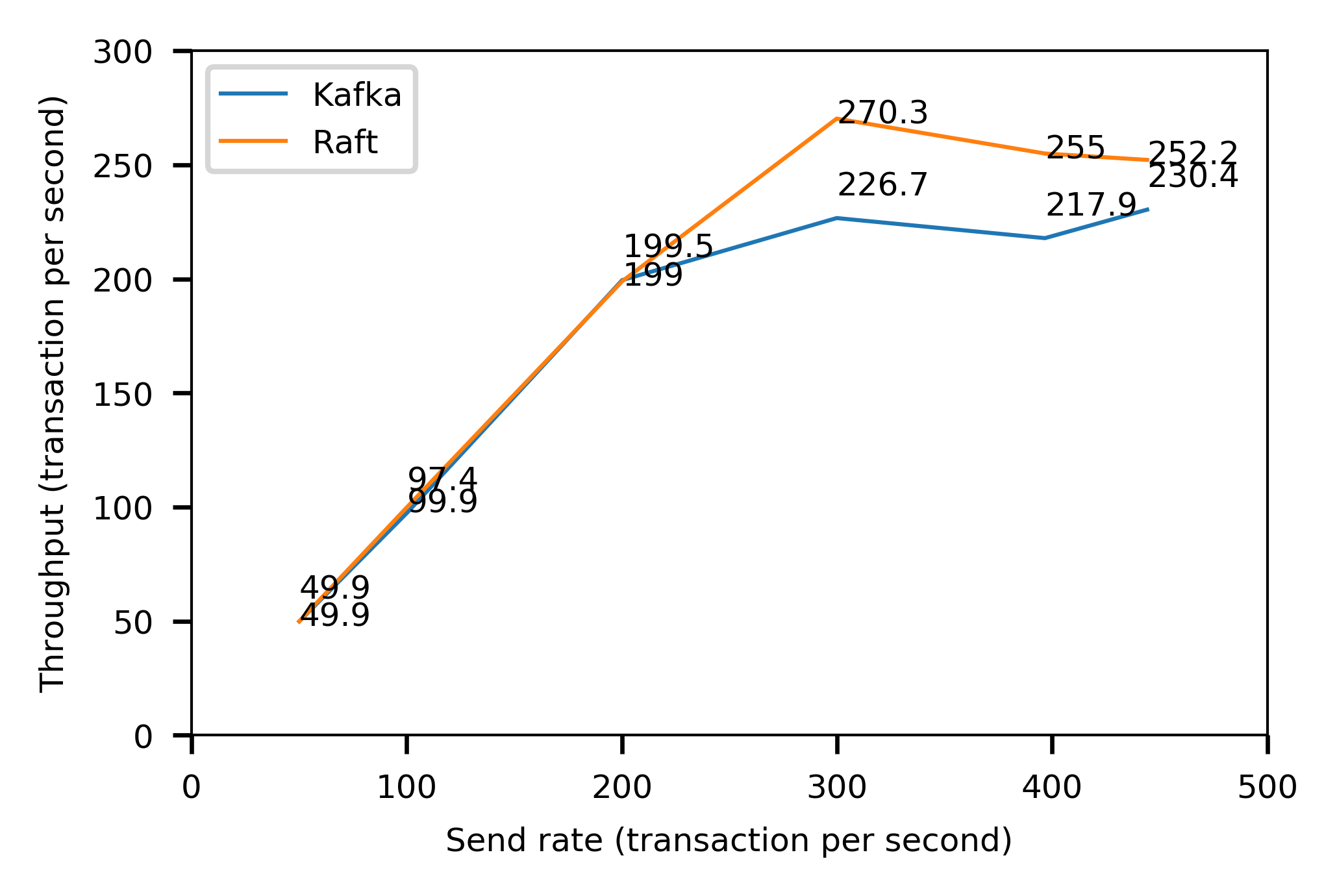}
    \caption{Throughput of Raft and Kafka under different transactions.}
    \label{fig:raft-kafka-TPUT}
\end{figure}

\Cref{fig:raft-kafka-latency,fig:raft-kafka-TPUT}  show the average latency and throughput for the policy decision transaction for Raft and Kafka based on different send rates. The number of transactions for these two tests is 5,000. The dendrite of 200 \ac{tps} is an optimal point for Kafka as it exhibits the lowest average latency,  and the average latency increases sharply after the send rate of 200 \ac{tps}.

The maximum throughput for both Raft and Kafka orderers is at the send rate point of 300 \ac{tps}; afterwards, the throughput drops for both of them. Overall, in terms of throughput and average latency, Raft performed better than Kafka when the throughput passed 200 \ac{tps} as a turning point for Kafka.

\begin{figure}
    \centering
    \includegraphics[width=1\columnwidth]{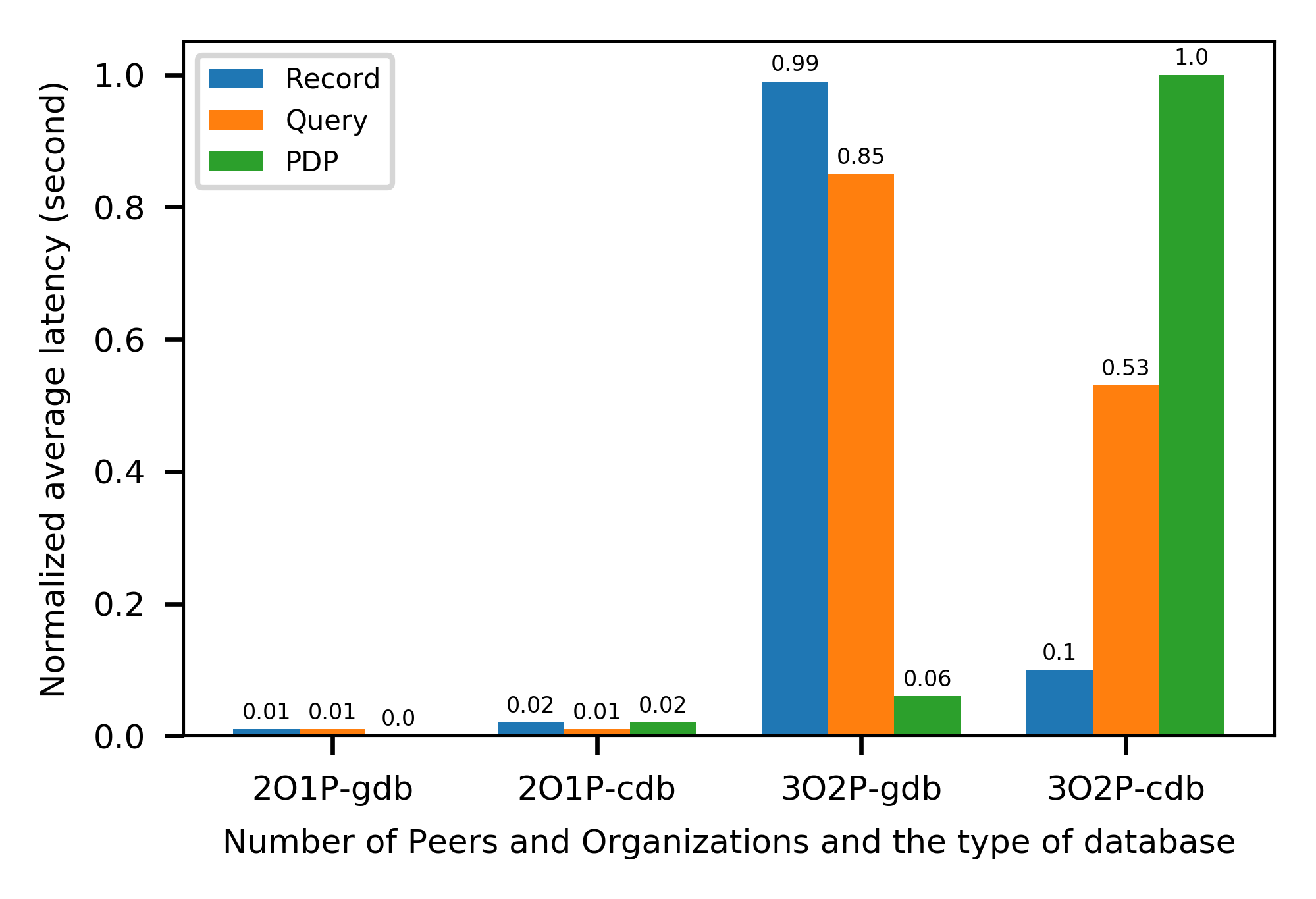}
    \caption{Throughput of Raft and Kafka under different transactions. Legend: $x$O$n$P = $x$ Organizations and $n$ Peers; gdb = GoLevelDB; cdb = CouchDB.}
    \label{fig:diffOrgandPeers}
\end{figure}


\begin{figure}
    \centering
    \includegraphics[width=1\columnwidth]{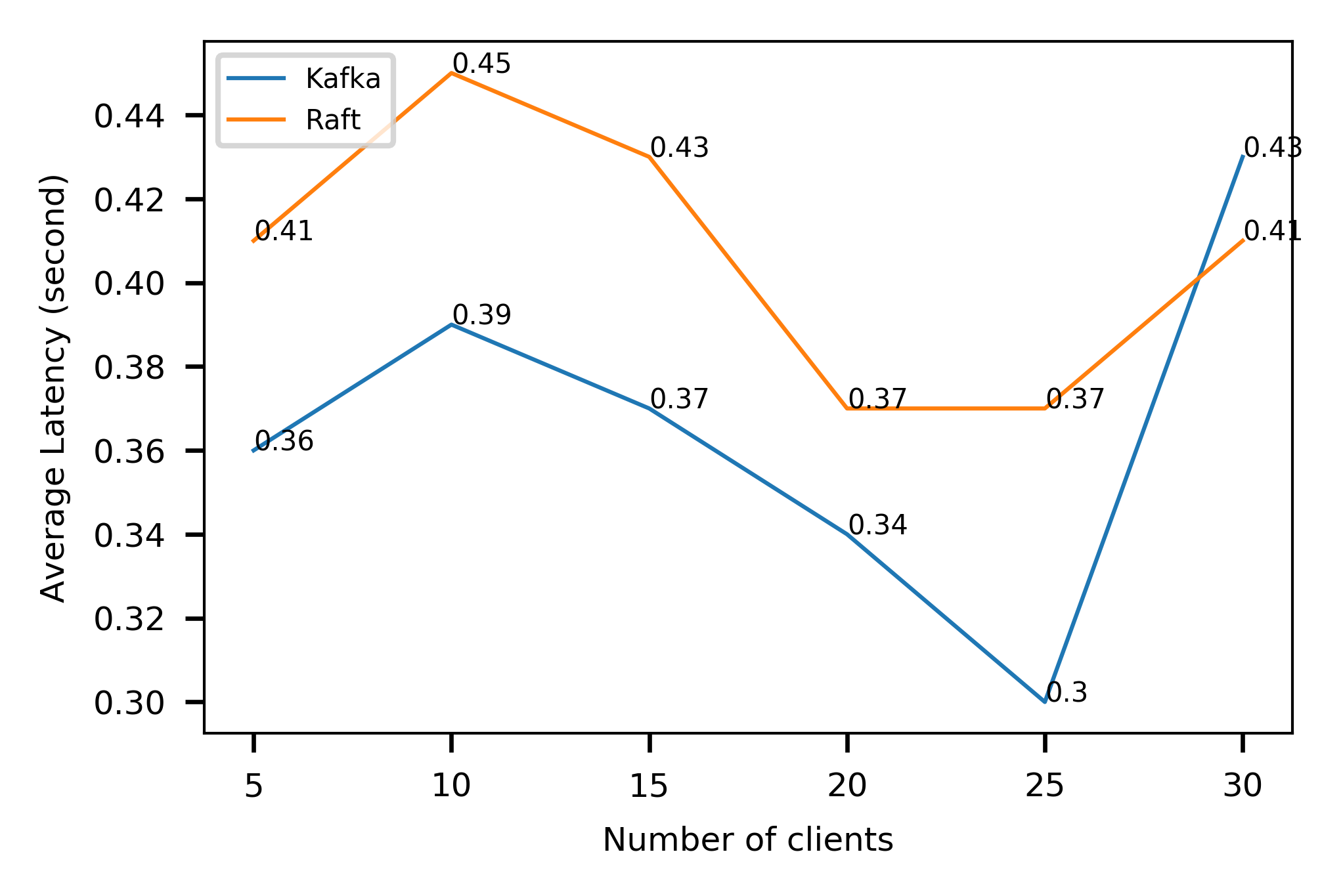}
    \caption{Average latency based of Raft and Kafka with different number of clients.}
    \label{fig:clients}
\end{figure}


\begin{table*}[]
\centering
\caption{Resource consumption for Raft and Kafka.}
\label{table:resourceConsumption}
\resizebox{\textwidth}{!}{%
\begin{tabular}{|c|c|c|c|c|c|c|c|c|c|}
\hline
Orderer          & Name           & Memory(max) & Memory(avg) & CPU(max) & CPU(avg) & Traffic In & Traffic Out & Disc Read & Disc Write \\ \hline
Raft             & dev-peer0.org1 & 74.4MB      & 71.MB       & 32.25\%  & 28.21\%  & 13.8MB     & 5.3MB       & 0B        & 0B         \\ \cline{2-10} 
                 & dev-peer0.org2 & 73.3MB      & 69.7MB      & 33.23\%  & 28.28\%  & 13.8MB     & 5.3MB       & 0B        & 0B         \\ \cline{2-10} 
                 & peer0.org1     & 379.3MB     & 369.4MB     & 67.21\%  & 56.32\%  & 31.1MB     & 23.7MB      & 0B        & 21.8MB     \\ \cline{2-10} 
                 & peer0.org2     & 284.0MB     & 274,1MB     & 70.78\%  & 55.66\%  & 31.1MB     & 23.6MB      & 4.0KM     & 21.8MB     \\ \cline{2-10} 
                 & orderer1       & 554.1MB     & 535.4MB     & 26.11\%  & 15.41\%  & 22.4MB     & 59.1MB      & 0B        & 37.2MB     \\ \cline{2-10} 
                 & orderer2       & 525.3MB     & 506.5MB     & 18.78\%  & 11.88\%  & 27.6MB     & 28.7MB      & 0B        & 37.0MB     \\ \cline{2-10} 
                 & orderer0       & 513.3MB     & 494.7MB     & 19.40\%  & 11.63\%  & 27.5MB     & 10.6MB      & 0B        & 37.2MB     \\ \hline
Kafka            & dev-peer0.org1 & 73.5MB      & 72.8MB      & 17.87\%  & 15.26\%  & 7.5MB      & 2.5MB       & 0B        & 0B         \\ \cline{2-10} 
                 & dev-peer0.org2 & 64.5MB      & 62.8MB      & 18.73\%  & 15.69\%  & 7.5MB      & 2.5MB       & 0B        & 0B         \\ \cline{2-10} 
                 & peer0.org1     & 295.9MB     & 286.2MB     & 52.08\%  & 49.15\%  & 27.1MB     & 17.0MB      & 368.0KB   & 21.2MB     \\ \cline{2-10} 
                 & peer0.org2     & 294.1MB     & 282.7MB     & 51.54\%  & 48.38\%  & 27.1MB     & 17.1MB      & 152.0KB   & 21.2MB     \\ \cline{2-10} 
                 & orderer0       & 121.1MB     & 113.1MB     & 25.80\%  & 23.30\%  & 29.6MB     & 11.1MB      & 4.0KB     & 18.4MB     \\ \cline{2-10} 
                 & orderer1       & 124.1MB     & 115.2MB     & 21.87\%  & 20.25\%  & 29.7MB     & 46.5MB      & 276.0KB   & 18.4MB     \\ \hline
\end{tabular}}
\end{table*}

\Cref{fig:diffOrgandPeers} shows the effect of increasing the number of organizations and peers and the comparison of two different databases, GoLevelDB and CouchDB. Increasing the number of organizations and peers increases the average latency for all three transactions. For the CouchDB database with three organizations and two peers, the average latency increases sharply to 43.81 seconds for the policy decision transaction, which is 17.73 times more than GolevelDB and 64.42 times more than the same database, with two organizations and one peer. It shows that CouchDB performs inadequately in comparison with GoLevelDB. 

\Cref{fig:clients} shows the average latency for both Raft and Kafka based on the different number of clients. This test is run based on 5,000 transactions and the policy decision transaction. For Kafka, 25 number of clients is like an optimal point that has the lowest average latency (0.3 seconds). However, 25 is an optimal point for the system with current resources. For Raft, as the graph shows, there are two points for the minimum average latency, corresponding to 20 and 25 number of clients. In general, Kafka performs better under 25 clients, but after 25 clients, it shows a sharp increase in the average latency.
Although it shows that the system is not scalable after 25 clients, it significantly depends on the computation power and limitations of the \ac{VM} instance. We have repeated the test with a more powerful \ac{VM} instance, and average latency was 0.36 second for Raft and 0.31 second for Kafka with 60 clients. 

\Cref{table:resourceConsumption} presents the resource consumption for policy decision transactions based on 5000 transactions for Kafka and Raft. As the presented numbers in the table demonstrate, Raft consumes 4.33 times more memory on average in comparison with Kafka. It clarifies that our early test with 10,000 transactions with Raft ordered failed because the \ac{VM} ran out of memory. 

\section{Conclusion}\label{sec:conclusion and future work}
Dependable accountability mechanisms are essential for audits.
In this paper, we discussed how permissioned blockchains could be helpful as a trustable backend in access control systems, thus providing a solid basis for audits. We proposed a  distributed  \ac{ABAC}  system based on Hyperledger Fabric with a focus on auditability and scalability. We validated our solution through a decentralized access control management application in digital libraries. First, we presented a comprehensive review of studies focusing on blockchain-based access control studies. Then we presented the system architecture and implementation details, where the \ac{PDP}, \ac{PAP}, and \ac{AM} components have been implemented using smart contracts on-chain, and the \ac{PEP} was implemented off-chain - based on the blockchain clients' requirements. 

The experimental evaluation of our solution considered various parameters based on the Hyperledger Caliper framework in terms of system performance. 
The evaluation results indicate that our system can effectively handle 10,000 
access request transactions with an average latency of 0.54 seconds.

Future work is in progress in two directions: first, building a robust framework and platform-independent solution towards distributed access control; second, integrating user authentication to our authorization solution.  
\section*{Acknowledgement} This research is supported by the Linux Foundation, in the context of the Hyperledger Fabric Based Access Control Project.

\bibliographystyle{spmpsci}      
\bibliography{bibliography}   


\begin{acronym}[BRBAC BN]
\acro{AAA}{Authentication, Authorization, and Accounting}
\acro{ABAC}{Attribute-Based Access Control}
\acro{AD}{Access Directory}
\acro{AM}{Attribute Manager}
\acro{API}{Application Programming Interface}
\acro{AP}{Attribute Provider}
\acro{AS}{Application Server}
\acro{BRBAC BN}{Blockchain Role-Based Access Control Business Network}
\acro{CC}{Cloud Computing}
\acro{CFT}{crash fault-tolerant}
\acro{DAC}{Discretionary Access Control}
\acro{DLT}{Distributed Ledger Technology}
\acro{EHR}{Electronic Health Record} 
\acro{EMR}{Electronic Medical Record}
\acro{ID}{Identifier}
\acro{IoT}{Internet of Things}
\acro{JSON}{JavaScript Object Notation}
\acro{MAC}{Mandatory Access Control}
\acro{PAP}{Policy Administration Point}
\acro{PDP}{Policy Decision Point}
\acro{PEP}{Policy Enforcement Point}
\acro{PIP}{Policy Information Point}
\acro{PKI}{Public Key Infrastructure}
\acro{PRP}{Policy Retrieval Point}
\acro{RBAC}{Role-Based Access Control}
\acro{SP}{Smart Policy}
\acro{TBAC}{transaction based access control}
\acro{VM}{Virtual Machine}
\acro{XACML}{eXtensible Access Control Markup Language}
\acro{XML}{Extensible Markup Language}
\acro{ZKP}{Zero knowledge proofs}
\acro{ftyp}{File Type}
\acro{tps}{transactions per second}
\end{acronym}

\end{document}